\documentclass[aps,prl,amssymb,amsmath,nobibnotes,nofootinbib,superscriptaddress]{revtex4}
\usepackage{amsfonts,amsmath,hyperref,url, color}
\usepackage{bm, bbm}
\usepackage{graphicx}
\usepackage{mathtools}
\usepackage{t1enc}
\usepackage{hepnames}

\newcommand{\bc}{\begin{center}}
\newcommand{\ec}{\end{center}}
\newcommand{\be}{\begin{eqnarray}}
\newcommand{\ee}{\end{eqnarray}}
\newcommand{\bs}{\begin{slide}}
\newcommand{\es}{\end{slide}}

\newcommand{\bi}{\begin{itemize}}
\newcommand{\ei}{\end{itemize}}

\begin{document}
\title{A bound on Planck-scale deformations of CPT from muon lifetime}

\author{Michele Arzano}
\affiliation{Dipartimento di Fisica ``E. Pancini" and INFN, Universit\`a di Napoli Federico II, Via Cinthia,
80126 Fuorigrotta, Napoli, Italy}

\author{Jerzy Kowalski-Glikman}
\affiliation{Institute for Theoretical Physics, University of Wroc\l{}aw, pl.\ M.\ Borna 9, 50-204
Wroc\l{}aw, Poland}
\affiliation{National Centre for Nuclear Research, ul. Pasteura 7, 02-093 Warsaw, Poland}

\author{Wojciech Wi\'{s}licki}
\affiliation{National Centre for Nuclear Research, ul. Pasteura 7, 02-093 Warsaw, Poland}

\date{\today}

\begin{abstract}
We show that deformed relativistic kinematics, expected to emerge in a flat-spacetime limit of quantum gravity, predicts different lifetimes for particles and their antiparticles. This phenomenon is a consequence of Planck-scale modifications of the action of discrete symmetries. In particular we focus on deformations of the action of CPT derived from the $\kappa$-Poincar\'e algebra, the most studied example of Planck-scale deformation of relativistic symmetries. Looking at lifetimes of muons and anti-muons we are able to derive an experimental bound on the deformation parameter of $\kappa \gtrsim 4\times 10^{14}$ GeV from measurements at the LHC. Such bound has the potential  to reach the value of $\kappa \gtrsim 2\times 10^{16}$ GeV using measurements at the planned Future Circular Collider (FCC).
\end{abstract}

\maketitle

Invariance under CPT, the combined action of charge conjugation, space inversion and time reversal, is a cornerstone of relativistic quantum mechanics and a fundamental symmetry of nature which has so far been left unquestioned by experiments \cite{Kostelecky:2008ts, Babusci:2013gda}. The CPT symmetry can be rigorously formulated in terms of a theorem which follows from the fundamental axioms of local quantum field theory \cite{Streater:1989vi}. Failure of any of these axioms would make room for potential violations of CPT invariance and thus any experimental evidence for violations of CPT symmetry could open the way to a radical re-thinking of the theoretical foundations of quantum field theory. On the other hand, conservation of CPT necessarily entails equality of masses and lifetimes of particles and antiparticles.

Quantum gravity effects have long been suggested as a potential source for violations of CPT.  As pointed out in ref. \cite{Wald:1980nm}, in the presence of a fundamental quantum-gravity-induced decoherence \cite{Hawking:1982dj}, the CPT operator is no longer well defined.  The possibility of experimentally testing CPT violation via decoherence effects has been widely explored (see e.g. refs. \cite{Huet:1994kr, Mavromatos:2004sz, Bernabeu:2006st}). As originally proposed in ref. \cite{Ellis:1983jz}, the neutral kaon system, thanks to the role played by quantum coherence over macroscopic distances, offers a particularly interesting phenomenological window to look for fundamental decoherence.

Violations of Lorentz invariance, possibly associated to discrete space-time or preferred frame features induced by quantum gravity (cf. e.g. ref. \cite{Kostelecky:1995qk}), are an alternative source for departures from the CPT invariance. Unlike the decoherence case mentioned above, for this type of departures from CPT one works within the usual quantum-field-theoretic setting with a well defined CPT operator and CPT violating terms arise in the effective low energy Lagrangian obtained from the specific quantum gravity model. As it turns out, such forms of CPT violation are remarkably constrained by current experimental data \cite{Kostelecky:2008ts}.

Planck-scale {\it deformations} of relativistic symmetries provide another scenario in which the ordinary notion of CPT symmetry must be re-thought. These models are linked to certain types of non-commutative spacetimes, widely studied in the past twenty years as candidates for a {\it Minkowski limit} of quantum gravity, and are based on momentum spaces with non-trivial geometric and algebraic structures described by a non-abelian Lie group. By far the most studied of these models is the $\kappa$-Poincar\'e algebra
\cite{Lukierski:1991pn,Lukierski:1992dt,Lukierski:1993wxa} related to the so-called $\kappa$-Minkowski non-commutative spacetime \cite{Lukierski:1993wx}, \cite{Majid:1994cy} (see ref. \cite{Kowalski-Glikman:2017ifs} for a recent review.)

As first realized in ref. \cite{KowalskiGlikman:2002ft} and later thoroughly explored in refs. \cite{KowalskiGlikman:2004tz, KowalskiGlikman:2003we}, the full structure of the $\kappa$-Poincar\'e algebra can be obtained in terms of the properties of a momentum space given by the Lie group $AN(3)$. Such group can be obtained from the Iwasawa decomposition of the five-dimensional Lorentz group $SO(4,1)\simeq SO(3,1)\, AN(3)$ \cite{KowalskiGlikman:2004tz} and this very structure ensures an action of the four-dimensional Lorentz group $SO(3,1)$ on the group manifold momentum space. Geometrically the $AN(3)$ group spans a submanifold of the four-dimensional de Sitter space $dS_4$ defined in an embedding five-dimensional Minkowski space as the subspace determined by the equation
\be
-p_0^2 + p_1^2 + p_2^2 + p_3^2 + p_4^2 =\kappa^2\, ;\,\,\,\,\,\,\,\, p_0+p_4>0\,.
\ee
Notice how the deformation parameter $\kappa$ is related to the curvature of the momentum manifold.

One can naturally associate translation generators\footnote{Choosing different parametrizations of $AN(3)$ leads, in general, to different choices of translation generators corresponding to different ``bases" of the $\kappa$-Poincar\'e algebra (See e.g. refs. \cite{Lukierski:1994ft,KowalskiGlikman:2002we,Arzano:2010jw} for a possible physical interpretation.)} $P_{\mu}$ to the
embedding coordinates $p_{\mu}$. This choice of translation generators is known as the ``classical basis" of the $\kappa$-Poincar\'e algebra \cite{Borowiec:2009vb} since the action of the Lorentz generators on $P_{\mu}$ and the corresponding coordinates $p_{\mu}$ is undeformed and the mass Casimir of the algebra is the usual one $P_0^2-{\mathbf{P}}^2= m^2$. In other words, at the algebra level the structures remain the same as in the ordinary Poincar\'e algebra, i.e.\ the generators of spacetime symmetries have the standard algebra of commutators. Also the way the generators act on one-particle states is standard. A full characterization of the symmetry generators, however, requires a specification of their action on multi-particle states, i.e. on tensor products of the one-particle ones. In the standard Poincar\'e case the action is Leibnizean reflecting the fact that the total momentum of a two-particles state is simply the sum of the momenta carried by its constituents. In the kinematics associated to the $\kappa$-Poincar\'e symmetries this is not the case and instead we find that the addition rule for spatial momenta and energy are given by
\begin{align}\label{sumk}
 \left(p^{(1)}\oplus p^{(2)}\right)_i &=
 p_i^{(1)}\frac{E^{(2)}+p_4^{(2)}}\kappa+p_i^{(2)}\,, \\ \left(E^{(1)}\oplus E^{(2)}\right) &=\frac1\kappa\, E^{(1)}\left(E^{(2)}+p_4^{(2)}\right) + \frac{\kappa E^{(2)}+\mathbf{p}^{(1)}\cdot \mathbf{p}^{(2)}}{E^{(1)}+p_4^{(1)}}
\end{align}
where
\begin{equation}\label{p4}
  p_4 = \sqrt{E^2-\mathbf{p}^2 +\kappa^2}\,.
\end{equation}
 This deformed composition of momenta is a direct consequence of the non-Abelian group law of $AN(3)$  (see ref. \cite{Kowalski-Glikman:2017ifs} for in-depth discussion.) In the limit $\kappa\rightarrow\infty$ the operation $\oplus$ becomes the ordinary summation: $\lim\limits_{\kappa \to \infty} \mathbf{p}^{(1)}\oplus \mathbf{p}^{(2)}= \mathbf{p}^{(1)}+ \mathbf{p}^{(2)}$.

There is a direct relation between the $\kappa$-deformed rule of momenta composition \eqref{sumk} and the noncommutative structure of spacetime, called $\kappa$-Minkowski space. The spacetime noncommutativity is encoded in the star product $\star$, which is used to multiply functions on spacetime. In particular, in the case of plane waves we have
\begin{equation}\label{starproduct}
  e^{i(E^{(1)}t-\mathbf{p}^{(1)}\mathbf{x})}\star e^{i(E^{(2)}t-\mathbf{p}^{(2)}\mathbf{x})}= e^{i((E^{(1)}\oplus E^{(2)})t-(\mathbf{p}^{(1)}\oplus\mathbf{p}^{(2)})\mathbf{x})}\,,
\end{equation}
with $\oplus$ defined in eq. \eqref{sumk}.

Similarly, the deformed inverse $\ominus$, defined by $(\ominus p)_i \oplus p_i=p_i \oplus(\ominus p)_i =0$, related to the $AN(3)$ group inversion law, is found to be
\begin{equation}\label{ominus}
  \ominus p _i  =-p_i \frac{\kappa }{E+p_4}\equiv S(p)_i\,.
\end{equation}
In the Hopf algebra terminology the deformed inverse is known under the name {\it antipode} and denoted by $S(p)_i\equiv \ominus p _i$, and we will use both symbols, interchangeably, in what follows. Notice that in the limit $\kappa\rightarrow\infty$, the operation $\ominus$ becomes the ordinary minus: $\lim\limits_{\kappa \to \infty} \ominus p_i = -p_i$.

In analogy to the undeformed case, for the $\kappa$-plane waves the antipode is associated with the operation of the $\kappa$-deformed conjugation $^\ddag$ \cite{Freidel:2007hk}, \cite{KowalskiGlikman:2009zu}.
\begin{equation}\label{dag}
  \left[e^{i(Et-\mathbf{p}\mathbf{x})}\right]^\ddag = e^{i(S(E)t-S(\mathbf{p})\mathbf{x})},
\end{equation}
so that
$$
e^{i(Et-\mathbf{p}\mathbf{x})}\star \left[e^{i(Et-\mathbf{p}\mathbf{x})}\right]^\ddag = \left[e^{i(Et-\mathbf{p}\mathbf{x})}\right]^\ddag\star e^{i(Et-\mathbf{p}\mathbf{x})}=1\,.
$$

The antipode is defined so as to preserve the form of the Casimir operator,
\begin{equation}\label{casacas}
E^2-\mathbf{p}^2=m^2 \quad \Leftrightarrow \quad S(E)^2 - S(\mathbf{p})^2 =m^2,
\end{equation}
where
\begin{equation}\label{sE}
  S(E) = \frac{\kappa^2}{E+p_4}-p_4  =  -E+\frac{\mathbf{p}^2}{E+p_4}\,.
\end{equation}
The energy of a particle with momentum $S(\mathbf{p})$ is equal to $-S(E) = \sqrt{m^2+S(\mathbf{p})^2}$.

Using these definitions we can now define the action of CPT operator $\Theta$ on plane waves\footnote{All $\Theta$-transformed states are multiplied by an arbitrary phase factor $e^{i\alpha}$ which hereon we consequently ignore as irrelevant.}. Following ref. \cite{Arzano:2016egk} we have
\begin{equation}\label{CPT1}
 \Theta\left( e^{iE(p) t - i\mathbf{p}\mathbf{x}} \right) = e^{-iS(E)(p) t - iS(\mathbf{p})\mathbf{x}}\,.
\end{equation}
The CPT operator $\Theta$ changes also all the charges to the opposite ones.

In what follows we will consider unstable particles. It is customary to effectively describe the particle decay process by adding to the energy an imaginary part $i\Gamma$ with $\Gamma = 1/\tau$ being the inverse of the state lifetime.
In order to implement this idea we start with the particle at rest, which is described by the wave function
\begin{equation}\label{decayrest}
\psi = \sqrt\Gamma\,  e^{-\Gamma t/2}\, e^{imt}\,.
\end{equation}
 We assume that similarly to the rest mass $m$, the decay rate $\Gamma$ is invariant under CPT transformation.
 We see therefore that the decay probabilities for particle and antiparticle are exactly the same
\begin{eqnarray}\label{probrest}
  {\cal P}_{\mbox{\scriptsize part}} & = & \psi\star\psi^\ddag \nonumber \\
              & = & \Gamma\, e^{-\Gamma t}    \nonumber \\
& = & \Theta\psi \star (\Theta\psi)^\ddag={\cal P}_{\mbox{\scriptsize apart}}.
\end{eqnarray}
It is important to notice a key property somehow obscured in the expressions for wave function \eqref{decayrest}  while apparent in eq. \eqref{probrest}. Namely, the time $t$ in \eqref{probrest} has the interpretation of the proper time that passed after the unstable particle was created. Therefore, strictly speaking it is the time interval and expressions are meaningful only for $t\geq0$.

In order to find out what is the decay probability of the particle in motion, we must make Lorentz transformation.
Here is the point when the expressions for particle and antiparticle start to differ. Namely it follows from the analysis presented in ref. \cite{Arzano:2016egk} that Lorentz transformation leaves invariant the mass shell relation $E^2-\mathbf{p}^2=m^2$ in the case of particle and $S(E)^2 - S(\mathbf{p})^2 = m^2$ in the case of the antiparticle. It follows that the $\gamma$ factor responsible for time dilation is $E/m$ in the case of the particle and
$$
-S(E)/m = \frac{E}m - \frac{\mathbf p^2}{\kappa m}
$$
in the case of the antiparticle.
Therefore the decay probability is equal to
\begin{equation}\label{decaypart}
  {\cal P}_{\mbox{\scriptsize part}} =  \frac{\Gamma E}m\,\exp\left(-\Gamma t\, \frac Em\right)
\end{equation}
for the particle and (in the leading $1/\kappa$ order)
\begin{equation}\label{decayapart}
  {\cal P}_{\mbox{\scriptsize apart}}  =\Gamma\left(\frac Em - \frac{\mathbf p^2}{\kappa m}\right) e^{-\Gamma t \left(\frac Em - \frac{\mathbf p^2}{\kappa m}\right)} 
\end{equation}
for antiparticle, where in both cases $t$ denotes the particle's proper time.

It should be noted that the deformation of CPT symmetry we propose here leads to a subtle violation of Lorentz symmetry. Indeed, the key point of our construction is a different form of Lorentz boost as applied to particles and antiparticles, from which the different expressions for decay rates follow. This effect certainly deserves further studies, especially with the connection of baryogenesis, as it might provide the mechanism for effective baryon-antibaryon asymmetry in early universe \cite{Sakharov:1967dj}.

Consequences of the $\kappa$-modification in eq. \eqref{decayapart} could  be examined experimentally by measuring precisely the lifetimes of particles and antiparticles, provided their energies are high enough to make the $\mathbf p^2$ at least partially compensating tiny effects of $1/\kappa$.
In addition, accuracy of the measurement of $\tau$ needs to be very good, for two reasons.
First, the correction $\mathbf p^2/(m\kappa)$ has to be comparable at least to experimental accuracy $\sigma_\tau/\tau$ that for known unstable particles amounts typically between $10^{-4}$ for mesons $\pi^{\pm}$ and $\PKzero$, and $10^{-6}$ for leptons $\mu^\pm$.
Second, any measurement of the lifetime requires experimentaly measured momenta to be Lorentz-transformed to the particle's rest frame.
Inaccuracies of laboratory momenta and energies thus propagate to the rest frame and, if large, can strongly affect $\sigma_\tau$.
This is particularly discouraging in non-accelerator experiments where energies of cosmic particles are occasionally very high, exceeding even $10^{6}$ GeV and thus $\mathbf p^2 \sim 10^{12}$ GeV$^2$ but, at the same time, experimental uncertainties usually tend to be large and hard to control.

In order to quantify our findings, in fig. 1 we plot the correction $\mathbf p^2/(\kappa m)$ for the muon which lifetime amounts to $\tau_\mu=(2.1969811\pm 0.0000022)\times 10^{-6}$~s \cite{PDG}.
As seen in fig. 1, if deformation $\kappa$ is close to the Planck mass $10^{19}$ GeV, as expected, any detectable correction requires momenta of the order $10^6$ GeV, unattainable at today's accelerating facilities.
Such energies are available in cosmic-ray experiments.
However, using them for our purposes would require a measurement of their lifetimes and reach very challenging accuracy of their energy determination.

Interesting enough, we can also estimate a limit on the deformation $\kappa$ that can be set for present energies at the Large Hadron Collider (LHC) and those planned at the Future Circular Collider (FCC) \cite{FCC}, both at CERN.
Using experimental accuracies of the lifetimes $\sigma_\tau$ and requiring $\mathbf p^2/(\kappa m)=\sigma_\tau$ for $\mathbf p=6.5$~TeV (LHC) and 50 TeV (FCC) one obtains the values of $\kappa$ labeling curves in fig. 1.
As can be seen there, the limiting value of $\kappa=4\times 10^{14}$ GeV can be obtained using muons at LHC and, in future, $\kappa=2\times 10^{16}$ GeV at FCC.
Any further improvement of the estimate of $\kappa$ at these energies requires progress in accuracy of the lifetime measurement.
But even using known experimental techniques there seems to be room for improvement since the most precise estimate of $\tau_\mu$ \cite{mulan} is still dominated by the statistical error.

\begin{figure}[h]
\includegraphics[scale=.5]{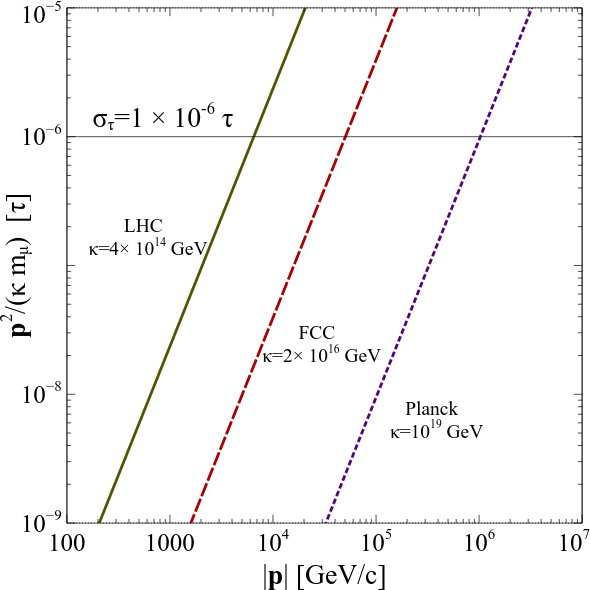}

\caption{Correction ${\mathbf p}^2/(\kappa m)$ to the muon lifetime. Horizontal line corresponds to the present experimental accuracy of the lifetimes \cite{PDG}. Two curves, labeled LHC and FCC, are for the deformation parameters $\kappa$ corresponding to corrections equal to experimental accuracies for maximal momenta at the Large Hadron Collider (continuous green, LHC) and the Future Circular Collider (dashed red, FCC). The violet dotted line corresponds to the Planck mass $\kappa = 10^{19}$ GeV.}
\end{figure}

Experimental setup for this kind of test is simple for muons.
The undeformed CPT operator $\Theta$ acts on the wave function of a muon of charge $q=\pm 1$ as follows
\begin{eqnarray}\label{thetamu}
\Theta\,\langle t,\mathbf x|q,E,\mathbf p\rangle & = & \langle -t,-\mathbf x|-q,E,\mathbf p\rangle^\ast \nonumber \\
                                                 & = & \langle t,\mathbf x|-q,E,\mathbf p\rangle,
\end{eqnarray}
where $^\ast$ stands for complex conjugation.
It is therefore sufficient to prepare a beam of monoenergetic muons propagating collinearly, measure times of their decays, identify their charges, and compare the lifetimes $\tau_{\mu^+}$ and $\tau_{\mu^-}$, as determined from slopes of spectra of their decay times in laboratory.
In practical terms, the $J/\psi$ resonance produced at very high energy and subsequently decaying into muons $J/\psi\rightarrow \mu^+\mu^-$ would be a useful tool for this study.

The method of studying subtle gravitational effects using decay laws (\ref{decaypart}, \ref{decayapart}) requires that all potential biases of the decay time distributions are precisely controlled.
According to studies initiated long time ago in ref. \cite{khalfin}, derivation of the decay law, taking into account a non-zero width of the mass distribution and integrating eq. (\ref{decayrest}) over mass $m$ with the Breit-Wigner distribution, leads to modification of the exponential function.
However, as studied by numerous authors, non-exponential terms appear only for large decay times exceeding tens of the mean lifetime and are orders of magnitude smaller than $\mathbf p^2/(\kappa m)$ (cf. ref. \cite{giacosa} and references therein).
The usual exponential formulas (\ref{probrest}) and (\ref{decaypart} \ref{decayapart}) are particularly accurate for particles where the ratio $\Gamma/m$ is small.
Since for the muon it amounts to $3\times 10^{-18}$, estimated corrections to the decay rate do not exceed $10^{-37} \Gamma$ \cite{giacosa} and therefore are by far smaller than those expected from gravitational effects considered in this paper.

\section*{Acknowledgments} MA acknowledges support from the COST Action  MP1405 "QSpace" for a Short Term Scientific Mission Grant supporting a visit to the University of Wroclaw where part of this work was carried out. The work of WW was partially supported by the Polish National Science Centre grant nr. 2017/26/M/ST2/00697,  while for JKG it is supported by Polish National Science Centre, projects number 2017/27/B/ST2/01902.

\end{document}